\documentclass[3p,times,procedia]{elsarticle}
\usepackage{nupha_ecrc}

\volume{00}

\firstpage{1}

\journalname{Nuclear Physics A}

\runauth{* et al.}

\jid{nupha}

\jnltitlelogo{Nuclear Physics A}

\usepackage{amssymb}
\biboptions{comma,square,sort&compress}

\usepackage[figuresright]{rotating}

\usepackage{nicefrac}
\usepackage{slashed}
\usepackage{amsmath,graphicx}
\usepackage[colorlinks=true,linktocpage=true,linkcolor=blue,citecolor=blue]{hyperref}
\usepackage{float}
\usepackage{nicefrac}
\usepackage[normalem]{ulem}
\usepackage{amsmath}
\usepackage{subcaption}

\begin{document}
%%%%%%%%%%%%%%%%%%%%%%%%%%%%%%%%%%%%%%%%%%%%%%%%%%%%%%%%%%%%%%%%%%%%%%%%%%
%%%%%%%%%%%%%%%%%%%%%%%%%%%%%%%%%%%%%%%%%%%%%%%%%%%%%%%%%%%%%%%%%%%%%%%%%%
\begin{frontmatter}
%%%%%%%%%%%%%%%%%%%%%%%%%%%%%%%%%%%%%%%%%%%%%%%%%%%%%%%%%%%%%%%%%%%%%%%%%%

\dochead{}

\title{Heavy Quark and Quarkonium Transport in High Energy Nuclear Collisions}

\author{Kai Zhou$^{1,2,3}$, Wei Dai$^1$, Nu Xu$^{4,5}$, and Pengfei Zhuang$^1$}

\address{$^1$ Physics Department, Tsinghua University and Collaborative Innovation Center of Quantum Matter, Beijing 100084, China\\
         $^2$ Institute for Theoretical Physics, Johann Wolfgang Goethe-University, Max-von-Laue-Str. 1, D-60438 Frankfurt am Main, Germany\\
         $^3$ Frankfurt Institute for Advanced Studies, Ruth-Moufang-Str. 1, D-60438 Frankfurt am Main, Germany\\
         $^4$ Nuclear Science Division, Lawrence Berkeley National Laboratory, Berkeley, CA 94720, USA\\
         $^5$ Key Laboratory of Quark and Lepton Physics (MOE) and Institute of Particle Physics, Central China Normal University, Wuhan 430079, China}

\begin{abstract}
The strong interaction between heavy quarks and the quark gluon plasma makes the open and hidden charm hadrons be sensitive probes of the deconfinement phase transition in high energy nuclear collisions. Both the cold and hot nuclear matter effects change with the colliding energy and significantly influence the heavy quark and charmonium yield and their transverse momentum distributions. The ratio of averaged quarkonium transverse
momentum square and the elliptic flow reveal the nature of the QCD medium created in heavy ion collisions at SPS, RHIC and LHC energies.
\end{abstract}


%%%%%%%%%%%%%%%%%%%%%%%%%%%%%%%%%%%%%%%%%%%%%%%%%%%%%%%%%%%%%%%%%%%%%%%%%%
\end{frontmatter}
%%%%%%%%%%%%%%%%%%%%%%%%%%%%%%%%%%%%%%%%%%%%%%%%%%%%%%%%%%%%%%%%%%%%%%%%%%

%%%%%%%%%%%%%%%%%%%%%%%%%%%%%%%%%%%%%%%%%%%%%%%%%%%%%%%%%%%%%%
\section{Introduction}
\label{s1}
%%%%%%%%%%%%%%%%%%%%%%%%%%%%%%%%%%%%%%%%%%%%%%%%%%%%%%%%%%%%%%
The strongly interacting Quantum Chromodynamics (QCD) system is expected
to experience a phase transition from hadronic matter to a new state of partonic matter, the so called quark-gluon plasma (QGP).
The existence of the partonic phase has been well established in $\sqrt{s_{NN}}$= 200 GeV Au+Au collisions at
RHIC as well as in $\sqrt{s_{NN}}$= 2.76 TeV Pb+Pb collisions at LHC. The study of high energy nuclear collisions
has entered a new era where current physics programs at both RHIC and LHC focus on the properties of the new form
of matter and how the hot matter undergoes the deconfinement transition. Since the new state of matter can exist only in the early stage of the collisions and cannot be directly observed in the final state, one needs sensitive probes to demonstrate the formation of the new state. Heavy flavor hadrons, including open and hidden charm or beauty hadrons, are widely considered as such probes. Different from light hadrons which are constructed from the constituents of the hot matter on the phase transition hypersurface and carry the collective properties of the matter, heavy quarks are mainly produced through hard processes in the initial
stage of the collisions, due to their large masses compared to the scale of QCD and the temperature of the medium. Then the initially produced quarkonia and the rest of the heavy quarks will pass through the whole medium and carry its properties in the final state.
By comparing the open charm production in A+A collisions with that in elementary p+p collisions, we can extract
information about the energy loss of heavy quarks
during their propagation through the hot medium. On the other hand, due to
color screening at finite temperature and density, quarkonium suppression in nuclear collisions has
been proposed to provide an unambiguous signature of QGP formation.

%%%%%%%%%%%%%%%%%%%%%%%%%%%%%%%%%%%%%%%%%%%%%%%%%%%%%%%%%%%%%%
\section{Heavy Quark Transport}
\label{s2}
%%%%%%%%%%%%%%%%%%%%%%%%%%%%%%%%%%%%%%%%%%%%%%%%%%%%%%%%%%%%%%
In high energy nuclear collisions, heavy quarks are investigated through measuring the decay products of open heavy mesons, like pions and kaons, or semi-leptonic
decay electrons and muons, or reconstructed non-prompt charmonia from $B$ decay. Due to the large mass, one would intuitively expect that, heavy quarks can hardly participate the medium collective motion and their thermalization should be difficult in comparison with light quarks. However, at both RHIC and LHC heavy quarks follow availably the flow
of the medium and possess a large elliptic flow $v_2$. This discovery indicates an extremely strong interaction between heavy quarks and the hot and dense medium, especially near the hadronization time when the collective flow of the bulk medium has already built up.
At the moment, what is the exact mechanism of the in-medium interaction of heavy quarks is still an open question and extensively studied
in both perturbative and nonperturbative approaches. For very hard heavy quarks with $p_T\gg m_Q$, the medium induced gluon radiation is expected to dominate the energy loss mechanism, since there should be no big difference between
heavy and light quarks in ultra-relativistic domain where quark mass is negligible. Due to the dead cone effect induced by the large
mass scale at small or intermediate $p_T$ range, the radiative process is reduced and the collisional energy loss
is invoked.

In the hot QCD medium, the typical momentum transfer in heavy quark interactions is of the order of the Debye mass $m_D\sim gT$. At the
current colliding energies, it is smaller than the heavy quark mass (strictly speaking, the thermal momentum), a Fokker-Planck approach
or Langevin approach is then suitable to treat the Brown motion of heavy quarks in medium, which can be deduced from the corresponding Boltzmann
equation without mean field interaction, see for instance Ref.~\cite{svetitsky} for details. In such stochastic treatment, the heavy quark in-medium interaction and the medium properties are encoded in the equation through the transport coefficients. In the Langevin approach, the three-momentum of an individual heavy quark evolves as
\begin{equation}
 \frac{d{\bf p}}{dt}=-\gamma(T,p){\bf p}+{\bf \eta}(t),
\end{equation}
where $\gamma$ is the drag coefficient giving the deterministic friction force, and $\bf \eta$ is a stochastic noise term
describing the random kicks from the medium constituents and satisfies
\begin{eqnarray}
 \eta_i(t)\eta_j(t')=\kappa(T,p)\delta_{ij}\delta(t-t')
\end{eqnarray}
in nonrelativistic case with $\kappa$ being the momentum-space diffusion coefficient (mean-squared momentum transfer per
unit time) or
\begin{eqnarray}
 \eta_i(t)\eta_j(t')=[\kappa_L \hat p^i \hat p^j +\kappa_T(\delta_{ij}-\hat p^i \hat p^j)]\delta(t-t')
\end{eqnarray}
in relativistic case with $\kappa_{L/T}$ being the
longitudinal and transverse momentum-space diffusion coefficients. One can decompose $\bf \eta$ into $\eta_L$ and $\eta_T$ representing random momentum kicks in
longitudinal and transverse direction with respect to the heavy quark motion.

In a generalized version of the Langevin approach, energy loss due to the medium-induced gluon radiation is included by adding a
recoil force~\cite{cao}
\begin{equation}
 \frac{d{\bf p}}{dt}=-\gamma(T,p){\bf p}+{\bf \eta}(t)+\bf \mathnormal{f}_g
\end{equation}
with ${\bf \mathnormal{f}_g}=-d{\bf p}_g/dt$, and the momentum of radiated gluons are sampled according to the gluon radiation spectrum
evaluated from pQCD higher twist calculation. To effectively introduce a balance between the gluon radiation and the inverse
absorption, a lower cut-off ($\pi T$) is set in the gluon radiation spectrum.

To ensure the thermal equilibrium in the limit of long time evolution, the drag coefficient $\gamma$
should be related to the momentum diffusion coefficient $\kappa$ (or $\kappa_{L/T}$ in relativistic case) through
generalized Einstein fluctuation-dissipation relation. In the case of weak coupling, the momentum
diffusion can be calculated in the frame of perturbative QCD at different levels, for instance at the leading order~\cite{svetitsky}, leading order with hard thermal loop resummation~\cite{moore,alberico}, next to leading order~\cite{huot}, and leading order with hard thermal loop resummation and running coupling~\cite{peigne,gossiaux,uphoff}. For nonperturbative diffusion coefficient in strong coupling, there are calculations from lattice QCD simulation~\cite{banerjee}, thermodynamic T-matrix
approach~\cite{riek}, AdS/CFT correspondence~\cite{gubser} and the semi-Quark-Gluon Monopole Plasma (sQGMP)
scheme~\cite{xu}. To describe simultaneously the nuclear modification factor $R_{AA}$ and elliptic flow $v_2$ of open heavy mesons,
the drag coefficient in the Langevin approach needs to have a strong enhancement around the critical temperature $T_c$. The sQGMP model
presents a consistent framework showing both soft bulk viscous and hard jet quenching with a smooth crossover from perturbative region
to nonperturbative region. The enhancement around $T_c$ for $\hat q/T^3$ and dropping off for $\eta/s$ are mainly due to the emergence of color-magnetic monopoles near $T_c$
which are the pertinent degrees of freedom with
long wavelength probably assuring the confinement transition. However, only the high $p_T$ heavy flavors can be described in
this framework. For the low $p_T$ part, the coalescence mechanism for a heavy quark to statistically pick up a light quark on the hadronization hyper surface can also contribute to the large
collective flow.

From the lattice QCD simulation~\cite{kaczmarek3}, there is a very large additional entropy around $T_c$ associated with
a heavy quark pair and it increases with the heavy quark distance in some range, reflecting an entanglement
between the heavy quark and the medium which induces numerous states for the heavy quark pair. Focusing on heavy quarks
(or considering long distance limit for heavy quark pairs), this indicates that, heavy quarks are strongly
coupled to the medium, especially near $T_c$. Phenomenologically, a hybrid treatment to separate the heavy quark evolution into two steps according to the temperature of the medium and take into account
the strong interaction near $T_c$ would be helpful to understand the heavy quark motion in QGP.

%%%%%%%%%%%%%%%%%%%%%%%%%%%%%%%%%%%%%%%%%%%%%%%%%%%%%%%%%%%%%%
\section{Heavy Quark Potential}
\label{s3}
%%%%%%%%%%%%%%%%%%%%%%%%%%%%%%%%%%%%%%%%%%%%%%%%%%%%%%%%%%%%%%
We now discuss the potential between a pair of heavy quark and anti-quark which is a traditional observable for understanding the QCD confinement. The often used non-relativistic Cornell potential~\cite{koma} in vacuum contains a Coulomb part ($\sim 1/r$) and a linear term ($\sim \sigma r$), the former is significant at small separation and the latter becomes dominant at large distance. The Cornell potential describes well the quarkonium spectroscopy in vacuum and thus provides a well-defined background to further test the medium effects at finite temperature and density. From the microscopic QCD theory itself, the Cornell potential can be understood
using the effective field theories NRQCD and pNRQCD in $1/m_Q$ expansion~\cite{barchielli}, where $m_Q$ is the heavy quark mass. When we put the heavy quark pair into a
QGP phase, the color screening will limit the range of the attractive binding force between the two quarks. When the screening radius $r_D$, which is inversely proportional to the temperature of the QGP, becomes smaller than the bound state radius, the pair will melt into the medium~\cite{matsui}. This inspires the sequential melting scenario that the quarkonium states with different binding energy have different melting temperature and disappear in the hot medium sequentially. The experiment data on $\Upsilon$ suppression at RHIC~\cite{star} and excited states at LHC~\cite{cms}
support the sequential melting scenario.

At finite temperature, based on the Euclidean correlator of the Polyakov loops, the static heavy
quark free energy $F(T,r)$ is calculated in lattice QCD in quenched~\cite{kaczmarek1} and (2+1)-flavor QCD case~\cite{kaczmarek2}, from which the internal energy
$U(T,r)$ can be derived through thermodynamic relations,
\begin{equation}
U(T,r)=F(T,r)-T\frac{\partial F}{\partial T}=F(T,r)+TS(T,r),
\end{equation}
where the difference between $F$ and $U$ is the entropy $S(T,r)=-\partial F/\partial T$ associated with the heavy quark pair.
The free energy from lattice simulation shows a saturation at large distance $r$ and finite temperature $T$, which is compatible with the
Debye screening expectation for heavy quark interaction. The question is what is the relation between the lattice calculated free energy and the heavy quark potential $V(T,r)$ which controls the quarkonium dissociation. By solving the Schroedinger equation for the heavy quark pair, the dissociation temperature for $J/\psi$ in the limit of $V=F$ is $T_{J/\psi}\simeq 1.2 T_c$ with $T_c$ being the critical temperature of the deconfinement phase transition, while in the other limit of $V=U$ $J/\psi$s can survive in the
deconfined plasma up to the temperature $T_{J/\psi}\simeq 2.1 T_c$, showing a much more strong binding. Phenomenologically, the comparison between a model calculation and the experimental data on excited $\Upsilon$ states is
sensitive to the choice of the heavy quark potential~\cite{liu}. On the other hand, based on the directly
calculated quarkonium correlator $G(r,\tau)$ from lattice QCD, the in-medium quarkonium spectrum can be evaluated
by the Maximum Entropy Method (with still a large uncertainty in the reconstruction), and the dissociation temperature can
be estimated from the disappearance of the peak of the spectral function. For $J/\psi$, the melting point is located at $T_{J/\psi}\simeq 1.5 T_c$ from the recent hot QCD simulation~\cite{ding}.

One should note that the above potential analysis assumes the medium to be stable and isotropic. However, in a realistic heavy ion collision,
due to the fast expansion and non-zero viscosity of the system, the produced medium would exhibit a local anisotropy in momentum space. Stronger
attraction is found to be aligned along the longitudinal direction in an anisotropic medium~\cite{dumitru},
which will enhance the quarkonium surviving temperature compared to the isotropic case. The relativistic correction
to the potential model will also enhance the $J/\psi$ dissociation temperature by $7\% - 13\%$ depending on the choice of the potential~\cite{guo}.

Apart from the screening of the attractive binding between heavy quarks, the bound states would also suffer a scattering
dissociation by the medium constituents, which is not captured by the above analysis and should manifest as a thermal
width in the spectral function. From thermal field theory, a static heavy quark potential at finite temperature is recently evaluated by
Fourier transformation of the hard thermal loop resumed gluon propagator~\cite{laine}. Due to the complexity of the
gluon self energy, an imaginary part is automatically developed for the heavy quark potential at $T\gg 1/r\sim m_D$,
\begin{equation}
V_{HTL}(r\sim\frac{1}{m_D})=-\frac{4}{3}\frac{\alpha_s}{r}(e^{-m_D r}+m_D)+i\frac{8}{3}\alpha_sT\int_0^{\infty}dt(1-\frac{sin(m_D r t)}{m_D r t})\frac{t}{(t^2+1)^2},
\end{equation}
where $m_D$ is the gluon Debye screening mass. This imaginary part smears the spectral peaks and leads to
a finite thermal width reflecting the thermal medium effects,
\begin{equation}
 \Gamma=\int \Psi^{\dag}[\text {Im}V]\Psi,
\end{equation}
with $\Psi$ being the in-medium charmonium wave function. The underlying physics of this finite width is interpreted as the Landau damping of the low frequency gauge field. In terms of the nonrelativistic field theory pNRQCD,
the heavy quark potential is evaluated, and an additional color singlet-to-octet thermal break-up mechanism is found to
contribute to the quarkonium thermal width~\cite{brambilla}, especially in the temperature range where $m_D$ is smaller than the binding
energy. Motivated by the effective field theory calculation, a lattice regularized QCD simulation based on NRQCD Lagrangian
is recently performed~\cite{burnier1}. The heavy quark potential is extracted from the simulated spectral function,
where the spectral peak and width are respectively related to the real and imaginary part of the potential. It shows
that, the imaginary part of the potential $\text{Im} V$ is compatible with the perturbative result~\cite{laine}, and
the real part $\text{Re} V$ from HotQCD collaboration with $N_f=2+1$ flavors is closer to the free energy $F$.
By parameterizing the potential in a generalized Gauss law approach and putting it into the Schroedinger equation, the
dissociation temperature for $J/\psi$ is $T_{J/\psi}\simeq 1.37 T_c$~\cite{burnier2}.

To understand the lattice QCD result from a point of view of many-body physics, a non-perturbative T-matrix approach is
developed~\cite{mannarelli}. In this approach, by fitting the lattice calculated free energy, the heavy quark
potential is recently evaluated~\cite{liu2} by taking a diagrammatic treatment through defining the potential within a
scattering equation. The extracted potential is found to be in between the free energy $F$ and the internal energy $U$ but very close
to $U$. Interestingly, the derivative of the complex potential, namely the binding force between the two heavy quarks, is stronger than both
from free energy and internal energy, implying a remnant confining string tension in the medium related with a long-range
correlation.

Since the produced QGP in heavy ion collisions expands rapidly and thus cools down fast, the potential analysis might not
be valid, because the heavy quark pair would hardly keep in a fixed eigenstate of the evolving Hamiltonian of the heavy quark system. Even further, whether a separate treatment for quarkonium evolution by taking the QGP only as a background really works is somehow questionable. Due to the large entropy associated to a quarkonium in medium observed in lattice QCD ~\cite{kaczmarek3}, especially near $T_c$, the entropic force~\cite{kharzeev} induced coupling between medium constituent and heavy quarks might be stronger than the binding between the two heavy quarks. This makes
the treatment of the heavy quark bound state as an color neutral entity unjustifiable, it should be influenced by the surrounding medium, since
each heavy quark in the bound state is strongly entangled with the medium. An entropic self-destruction scenario is proposed and results in a stronger dissociation around the critical temperature $T_c$. Taking into account the entropic force, the quarkonium binding potential is argued to
be the free energy $F$~\cite{satz}, since the extra binding through the difference between $F$ and $U$ is cancelled by the repulsive entropic force.

%%%%%%%%%%%%%%%%%%%%%%%%%%%%%%%%%%%%%%%%%%%%%%%%%%%%%%%%%%%%%%
\section{Nuclear Matter Effects on Quarkonia}
\label{s4}
%%%%%%%%%%%%%%%%%%%%%%%%%%%%%%%%%%%%%%%%%%%%%%%%%%%%%%%%%%%%%%
Quarkonium suppression has been extensively studied both experimentally and theoretically since the first work by Matsui and Satz~\cite{matsui}.
From the experimental data and the corresponding theoretical interpretations, it is now realized that, apart from the color screening, some other medium
effects come into play during the quarkonium evolution which make the picture for charmonium production in heavy ion collisions complicated. They
are usually classified into cold and hot nuclear matter effects.

Before the initially produced quarkonia pass through the hot medium, they are already affected by the cold nuclear matter effects, including (1) nuclear absorption or normal suppression,
namely that the quarkonia are eaten up by collisions with the surrounding primary nucleons~\cite{abreu}; (2) Cronin effect, namely that prior to any hard scattering the initial partons acquire additional transverse momentum via multi-scattering with nucleons, and then the $p_T$ broadening will be inherited
by the produced quarkonia through hard processes~\cite{cronin}; (3) nuclear shadowing effect, namely that the parton distribution function (PDF) in nucleus differs from that in a free nucleon, and the suppressed or enhanced parton distributions, depending on the momentum fraction of the partons, will change the initial quarkonium production~\cite{mueller,vogt}. A deep understanding of the parton distributions in nuclei requires the knowledge from e+p and e+A collisions~\cite{frankfurt}.
These cold nuclear matter effects are expected to be fixed through p+A collisions, under the assumption that no hot medium is created in such collisions.

After entering the created partonic medium, the color screening will alter the binding inside the bound state.
Different suppression mechanisms work here in different temperature domains. When the temperature is
higher than the charmonium dissociation temperature, no bound states survive, all the quarkonium states melt into the medium. Below this melting threshold,
there is still charmonium suppression through collisions with the medium constituents like gluon dissociation~\cite{bhanot} and quasi-free dissociation~\cite{grandchamp}, which respectively correspond to singlet-to-octet transition and Landau damping in terms
of effective field theory. Microscopically, knowing the parton induced break-up cross section $\sigma_{\Psi}$,
the dissociation rate or thermal width can be expressed as
\begin{equation}
\Gamma_{\Psi}=\int\frac{d^3{\bf k}}{(2\pi)^3}v_{rel}\sigma_\Psi(s,T)f_g({\bf k}),
\end{equation}
for gluon dissociation and
\begin{equation}
\Gamma_{\Psi}=\sum_{i=g,q,\bar q}\int\frac{d^3{\bf k}}{(2\pi)^3}v_{rel}\sigma_\Psi(s,T)f_i({\bf k})(1\pm f_i({\bf k})),
\end{equation}
for quasi-free dissociation with $f_{p,g}({\bf k})$ the parton thermal distribution in medium. Inversely, this provides
a way to extract the microscopic dissociation cross section from the dissociation rate~\cite{brambilla,escobedo,brambilla2}.

Another hot medium effect related to the quarkonium suppression is the quarkonium regeneration inside the hot medium~\cite{pbm,thews,rapp,yan} through the recombination of those uncorrelated heavy quarks. With increasing colliding energy, the suppression of the initially produced quarkonia becomes more and more strong, but the regeneration contribution increases due to the abundant heavy quarks in the medium at high energies. The balance between the suppression and the regeneration controls the quarkonium production in high energy nuclear collisions. There are different ways to treat the suppression and regeneration. Within the statistical
hadronization model~\cite{pbm}, the quarkonia are statistically produced at the
hadronization transition from kinetically equilibrated medium, and all the initially produced quarkonia are assumed to be absorbed by the medium. Considering the fact that
$J/\psi$ and $\Upsilon$ can survive in the medium with temperature above the critical temperature $T_c$ and below the dissociation temperature $T_{J/\psi} (T_\Upsilon$), a continuous regeneration
scheme is proposed in kinetic formalism~\cite{thews} where quarkonia can be both dissociated
and regenerated continuously inside the QGP medium. In a strongly coupled QGP phase, the charm quark diffusion constant is very small, and the binding force between the two quarks inside a charmonium is still strong~\cite{shuryak}, especially near $T_c$,  a generalized Langevin approach with additional binding force (${\bf F}=-\nabla V$) is proposed
to evaluate the charmonium evolution, where the regeneration from correlated charm quarks is included.
More generally, being motivated by the fact that the QGP produced in heavy ion collisions is a rapid cooling system, open quantum
system formalisms are developed for studying quarkonia in nuclear collisions~\cite{akamatsu} which exclude the adiabatic condition used in usual treatments.
Similar to the open quantum system approaches, a Hamiltonian based
approach is developed, and through a chain of well-controlled approximations it finally arrives at a generalized
Langevin formalism, with which both the quarkonium dissociation and regeneration are calculable~\cite{blaizot}.

%%%%%%%%%%%%%%%%%%%%%%%%%%%%%%%%%%%%%%%%%%%%%%%%%%%%%%%%%%%%%%
\section{Quarkonium Transverse Momentum Distribution}
\label{s5}
%%%%%%%%%%%%%%%%%%%%%%%%%%%%%%%%%%%%%%%%%%%%%%%%%%%%%%%%%%%%%%
Transverse motion is developed during the dynamical evolution of
the system. It has been well documented in the light quark sectors
for heavy ion collisions at all energies. In
order to understand the production and suppression mechanisms and
extract the properties of the medium, it is proposed to construct a new
ratio for the second moment of the transverse momentum distribution.
The ratio $r_{AA}$ is defined as~\cite{zhou1,zhou2}
\begin{equation}
r_{AA}=\frac{\langle p_t^2 \rangle_{AA}}{\langle p_t^2
\rangle_{pp}}.
\end{equation}

\begin{center}
\begin{figure}[h]
\includegraphics[width=0.6\textwidth]{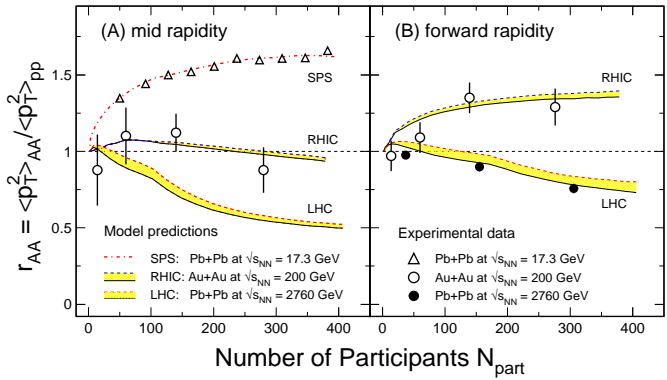}
\caption{(color online)  $J/\psi\ r_{AA}=\langle p_t^2\rangle_{AA}/\langle
p_t^2\rangle_{pp}$ as a function of collision centrality ($N_{part}$)
for collisions at SPS (open triangles)\cite{spsptraa}, RHIC (open
circles)\cite{rhic2007} and LHC (filled circles)\cite{aliceptraa}.
Left and right plot are results from mid-rapidity and
forward-rapidity, respectively.  The solid and dashed lines indicate
the model results with and without considering the shadowing
effect~\cite{eks}. }
\label{fig1}
\end{figure}
\end{center}

The quarkonium motion in heavy ion collisions can be described by a transport equation~\cite{zhu,yan} with initial condition determined by the superposition of p+p collisions and modification from the cold nuclear matter effects. The loss and gain terms on the right hand side of the transport equation are controlled by the quarkonium suppression and generation~\cite{liu3,tang}. Fig.\ref{fig1} shows the calculated $J/\psi$ $r_{AA}$~\cite{zhou2} as a function of collision
centrality ($N_{part}$) for collisions at SPS, RHIC and
LHC. There is a clear rise in
$r_{AA}$ in peripheral collisions at all colliding energies. It
is believed due to the Cronin effect before
the quarkonium formation~\cite{cronin} and it is more
important for collisions at lower energy.
%%%%%%%%%%%%%%%%%%%%%%%%%%%%%%%%%%%%%%%%%%%%%%%%%%%%%%%%%%%%%%%%%%%%%%%%%%%%
%%%%%%%%%%%%%%%%%%%%%%%%%%%%%%%%%%%%%%%%%%%%%%%%%%%%%%%%%%%%%%%%%%%%%%%%%%%%%%%

The centrality dependence of $r_{AA}$ clearly reflects the
underlying $J/\psi$ production and suppression mechanisms in high
energy nuclear collisions. At lower collision energy where the charm
cross section is small and the regeneration is negligible, almost
all of the observed $J/\psi$s are produced at the initial impact. In
this case, the Cronin
effect~\cite{cronin} is dominant and tends to increase the
transverse momenta of the finally observed quarkonia. As a result
$r_{AA}$ is always above unity and increases monotonically versus
collision centrality, see left panel of Fig.~\ref{fig1}. In the
truly high energy heavy ion collisions, on the other hand, charm
quarks are copiously produced and the regeneration for quarkonia can be significant. Although these heavy quarks
are initially produced via hard process with large transverse
momenta, they loss energy when interacting with the medium.
Since the temperature is high in such violent collisions,
especially in the most central collisions, almost all initially
produced $J/\psi$s are destroyed and the regenerated $J/\psi$s
dominate the final production. As shown in the left panel of
Fig.\ref{fig1}, the predicted $r_{AA}$ for heavy ion collisions at
LHC is less than unity and decreases toward more central collisions.
The prediction has been confirmed by the experimental results at
somewhat forward rapidity window as shown in the right plot of
Fig.\ref{fig1}~\cite{aliceptraa}. The strong competition between the
initial parton scattering and the regeneration at RHIC energy leads
to a flat centrality dependence of $r_{AA}$ at mid-rapidity,
shown as open circles in the left plot. Shift a bit away from the
mid rapidity, see the right plot of Fig.\ref{fig1}, the ratio at
RHIC is increasing versus centrality signaling the importance of
initial scattering at the RHIC energy. One may ask oneself if the
above result of $r_{AA}$ is due to the initial shadowing effect
which changes the parton distributions in nuclei. Different from the integrated
yield, the averaged transverse momentum is a normalized quantity,
therefore the shadowing induced change in the parton
distribution is minimized. As shown in Fig.\ref{fig1},  the small difference between the
solid and dashed lines, the hatched band, is the results of shadowing effect~\cite{eks}.

The elliptic flow is a measure of the collective motion and sensitive to the thermalization of the quarkonia. It describes the asymmetry of the quarkonium transverse motion,
\begin{eqnarray}
 v_2=\langle \frac{p^2_x-p^2_y}{p^2_x+p^2_y}\rangle = \langle \cos 2\phi \rangle.
\end{eqnarray}
The initially produced $J/\psi$s only
possess a very small elliptic flow from the geometry induced leakage effect~\cite{leakage}. The regeneration
on the other hand depends much on the heavy quark evolution. The large $v_2$ of charm quarks due to the strong interaction with the medium can be inherited by the regenerated $J/\psi$s. The transport approach can explain~\cite{liu4,zhou2} well the disappeared elliptic flow at RHIC energy~\cite{flow1} and the sizeable elliptic flow at LHC energy~\cite{flow2}.

%%%%%%%%%%%%%%%%%%%%%%%%%%%%%%%%%%%%%%%%%%%%%%%%%%%%%%%%%%%%%%
\section{Conclusion}
\label{s6}
%%%%%%%%%%%%%%%%%%%%%%%%%%%%%%%%%%%%%%%%%%%%%%%%%%%%%%%%%%%%%%
Heavy quarks and quarkonia are sensitive probes of the quark gluon plasma created in high energy nuclear collisions, due to their strong interaction with the constituents of the hot medium.
By taking into account the cold and hot nuclear matter effects on their evolution in the partonic and hadronic phases, we can describe most of the experimental data by adjusting some parameters in phenological models. However, there exist still some puzzles which are beyond our expectations, like the simultaneous treatment of the heavy quark $R_{AA}$ and $v_2$, the double ratio of $\psi'$ to $J/\psi$ yield in some special rapidity, the strong charmonium enhancement in peripheral collisions, and the partial thermalization of charmonia in small systems.
\noindent {\bf Acknowledgement:}
The work is supported by the NSFC and MOST grant Nos. 11335005, 11575093, 2013CB922000 and 2014CB845400, and the Helmholtz
International Center for FAIR within the framework of the LOEWE program launched by the State of Hesse.

%%%%%%%%%%%%%%%%%%%%%%%%%%%%%%%%%%%%%%%%%%%%%%%%%%%%%%%%%%%%%%%%%%%%%%%%%%
\end{document}